\documentstyle[12pt,aaspp4,flushrt,psfig]{article}
\begin{document}

\def\bq{\begin{equation}}
\def\eq{\end{equation}}
\def\bqy{\begin{eqnarray}}
\def\eqy{\end{eqnarray}}
\def\bqyn{\begin{eqnarray*}}
\def\eqyn{\end{eqnarray*}}
\def\bc{\begin{center}}
\def\ec{\end{center}}

%

\title{HARD X-RAY OBSERVATION OF ABELL 496}

\author{
Azita Valinia\altaffilmark{1,2}, Keith Arnaud\altaffilmark{1,2}, 
Michael Loewenstein\altaffilmark{1,2},   
Richard F. Mushotzky\altaffilmark{1}, and Richard Kelley\altaffilmark{1}
}

\altaffiltext{1}{Laboratory for High Energy
Astrophysics, Code 662, NASA's Goddard
Space Flight Center, Greenbelt, MD 20771; valinia@milkyway.gsfc.nasa.gov}
\altaffiltext{2}{Department of Astronomy, University of Maryland,
College Park, MD 20742}

\bc
Accepted for Publication in the Astrophysical Journal
\ec

\begin{abstract}
We report the results of hard X-ray observations of Abell 496 (A496),
a nearby relaxed cluster, using the {\it Rossi X-ray Timing Explorer}
(RXTE).  The 3-20~keV spectrum of this cluster is well-modeled by a
thermal component of $kT \sim 4.1$~keV plus a cooling flow with mass
accretion rate of $\dot{M} \sim 285 \, {\rm M_{\odot} \, yr^{-1}}$.
The spectrum is equally well-modeled by a
multi-temperature plasma component with a Gaussian temperature distribution
of mean temperature 
3.8~keV and $\sigma_{kT} \sim
0.9$~keV. The metallicity is found to be approximately 1/3 solar; however, 
the Ni/Fe ratio is about 3.6.  No
significant nonthermal emission at hard X-rays was detected for this
cluster. We discuss the implications of the models presented here and
compare them with the temperature profiles derived for this cluster
using the {\it Advanced Satellite for Cosmology and Astrophysics}
(ASCA). Our results are inconsistent with declining 
temperature profiles. 
\end{abstract} 
 

\newpage
%
 
\section{INTRODUCTION}

Observations over the last few years using ASCA and BeppoSAX have 
revolutionized our understanding of the X-ray emission from clusters
of galaxies. Spatially-resolved spectroscopy has led to measurements
of temperature and abundance variations in the hot gas trapped in
the cluster gravitational wells. These measurements can be used to
derive the total mass of clusters, to look for the signatures of
merger events, and to follow the chemical evolution of the gas and
the history of star formation in the cluster.

However, these results have not been without their controversies.
Markevitch et al. (1998; 1999) have claimed, using ASCA observations,
that most clusters have temperatures decreasing with distance
from the cluster center. This analysis has been disputed using the
same ASCA data by White (2000), using ROSAT PSPC data by Irwin,
Bregman \& Evrard (1999), and using BeppoSAX data by Irwin \& Bregman 
(2000). Determining which of these analyses is correct is important
for determining the true gravitational mass profiles in clusters of
galaxies and for comparison with cosmological simulations.

In addition, there is a lively debate about the significance
of any extended non-thermal emission from clusters of galaxies. This
is claimed to be detected both at low energies (e.g. Lieu et al. 1996;
Kaastra et al. 1999) and at high energies (e.g Fusco-Femiano et
al. 1999; Rephaeli, Gruber \& Blanco 1999; Kaastra et al. 1999;
Kaastra 2000). Such non-thermal emission could be due to
inverse-Compton emission from cosmic microwave background photons scattering
off relativistic electrons. If these electrons are also detected in
the radio then the magnetic field strength can be determined. 
Alternatively, the emission could be due to non-thermal bremsstrahlung
from suprathermal electrons accelerated in shocks or turbulence.
However, Arabadjis \& Bregman (1999) found no evidence
for an extremely soft non-thermal X-ray emission in a sample of 
clusters of galaxies.

Finally, there is a dispute about whether the Fe K$\alpha$ resonance
line is optically thick to resonance scattering in the cores of
clusters. If it were, this would mean that Fe abundances have been 
systematically underestimated. The detection of resonance scattering
hinges on comparing the strengths of the Fe K$\alpha$ and Fe K$\beta$
complexes. Since Fe K$\beta$ is confused with Ni K$\alpha$ using
current spectrometers the disagreement comes down to the Ni abundance
assumed (Akimoto et al. 1997; Molendi et al. 1998; Dupke \& Arnaud
2000). The Fe/Ni ratio is a sensitive indicator of the types of 
supernovae whose products have enriched the gas (Dupke \& White 2000).
Our understanding of the chemical evolution of clusters depends on this
issue being resolved.

We will not answer all these questions with an RXTE observation of a
single cluster. However, we will argue in this paper that high signal
to noise (S/N) 
spectra from a large beam but well-calibrated detector such as the PCA 
can impose important constraints on models for the cluster emission. 
Previous high S/N single beam observations of selected clusters have been 
done by Johnstone et al. (1992) using Ginga and by Henriksen \& White (1996)
using HEAO-1. Since RXTE 
has no spatial resolution within the cluster it cannot be used to
determine the temperature variation with position but it can test
models proposed on the basis of observations with other satellites. 
The bandpass of the PCA extends above 20 keV, making it sensitive
to non-thermal emission, particularly when observing cooler clusters.
Finally, an observation of the entire cluster means that any effect
due to (isotropic) resonance scattering will be absent.

\section{THE CLUSTER A496}

We chose to observe A496, a nearby ($z \approx 0.033$) cluster
with a moderate cooling flow. Recent optical and X-ray studies of this
cluster (e.g. Markevitch et al. 1999; Durret et al. 2000) indicate
that it has a regular morphology and a small amount of substructure.
The analysis of the position, luminosity and velocity dispersion of
the resident galaxies indicate that it is in a relaxed state with no
strong environmental effects. The study of this cluster is, therefore,
ideal as a prototype for understanding the properties of galaxy
clusters. In addition, its relatively low X-ray temperature of 4--5 keV
makes it an excellent prospect for testing for the presence of
a high energy non-thermal tail to the spectrum.

Early observations of A496 with the Einstein SSS and Ginga LAC imply
that metal abundances are centrally enhanced (White et al.
1994). Mushotzky et al. (1996) determined elemental abundances of O,
Ne, Mg, Si, S, Ca, Ar, and Fe from ASCA data and concluded that the
observed ratio of the relative abundances of elements outside of the
central cooling flow region is consistent with an origin of all the
metals in Type II Supernovae. Dupke \& White (2000) confirm central
iron abundance enhancements but also find a central nickel-to-iron
abundance ratio of approximately 4 that is consistent with the
``convective deflagration'' model in SN Ia explosion models.

Markevitch et al. (1999) derive maps and radial profiles of the gas
temperature from ASCA measurements and find an average cluster
temperature of $4.7\pm 0.2$~keV (90\% error), about 10\% higher
than the wide beam single-temperature fits. Outside the central
cooling flow region and within ASCA's field of view (fov), they
describe the temperature profiles by a polytropic gas model with
$\gamma \sim 1.24$. On the other hand, White (2000) finds the temperature
outside of the cooling region to be isothermal. Dupke \& White (2000) also
find a flat temperature profile outside the cooling radius. 
Kaastra (2000) notes that BeppoSAX observations 
of A496 suggest a hard tail in the spectrum. Irwin \& Bregman (2000) 
find a temperature rising from just below 4 keV in the center 
to approximately 5 keV at a radius of 9 arcminutes. 

\section{OBSERVATIONS} 

A496 was observed with the PCA and HEXTE instruments on board {\it
RXTE} during December of 1998 and January of 1999 for a total duration
of 100~ks.  The PCA (Jahoda et al. 1996) has a total collecting area
of 6500~${\rm cm^2}$, an energy range of $2-60$~keV, and energy
resolution of $\sim 18\%$ at 6~keV. The collimator field of view is
approximately circular ($2^\circ$ diameter) with FWHM of
$1^\circ$. The HEXTE (Rothschild et al. 1998) consists of two
clusters, each having a collecting area of 800~${\rm cm^2}$, an energy
range of $15-250$~keV, energy resolution of 15\% at 60~keV, and a
field of view of $1^\circ$ FWHM. Furthermore, each cluster ``rocks''
along mutually orthogonal directions to provide background
measurements away from the source.

We used the data taken when all 5 detectors of the PCA were on and the
elevation angle from the limb of the Earth was greater than
$10^\circ$.  For this time duration, the background subtracted count
rate over the $3-20$~keV band was $24.4\pm0.0279\,{\rm
counts\,s^{-1}}$ for a total of more than 2.4 million counts.  
The total on-source integration time for the HEXTE
detectors was $\sim 30$~ks.  The background subtracted count rate over
the $15-30$~keV energy band was $0.00534\pm0.0313\,{\rm
counts\,s^{-1}}$ and $-0.0259\pm0.0259\, {\rm counts\,s^{-1}}$ for
HEXTE clusters A and B, respectively.  Hence, there was no positive
detection with HEXTE.  In what follows, we present results of our
analysis of the PCA data.

\section{RESULTS}  

\subsection{Reliability of the PCA calibration}

Our conclusions depend critically on the accuracy of the PCA response
matrix used to fit the A496 data. Because the statistical accuracy
of our data is so high we have to be particularly careful with
systematic effects. We have analyzed observations of the Crab
nebula obtained on December 15 and 29 of 1998 and January 13 of 1999, the
same months that A496 was observed. We used pcarsp v2.43 to make the
PCA response. We fit all three observations simultaneously over the
3--15~keV band. 
Figure 1 shows the ratio of the data to model for the best fit
absorbed power-law, which has an average slope of 2.167 and an average
normalization of 7.3 ${\rm
photons \, keV^{-1} \, cm^2 \, s^{-1}}$ at 1~keV.
The best fit column density was $1.7 \times 10^{21} \, {\rm cm^{-2}}$.
The remaining systematic errors are at the 1\% level or smaller. In the
following description of our analysis of A496 we show ratios of data
to model and assume that any variations above those seen in the Crab
represent real differences between the assumed model and the data.

\begin{figure}[h]
\centerline{
{\hfil
\psfig{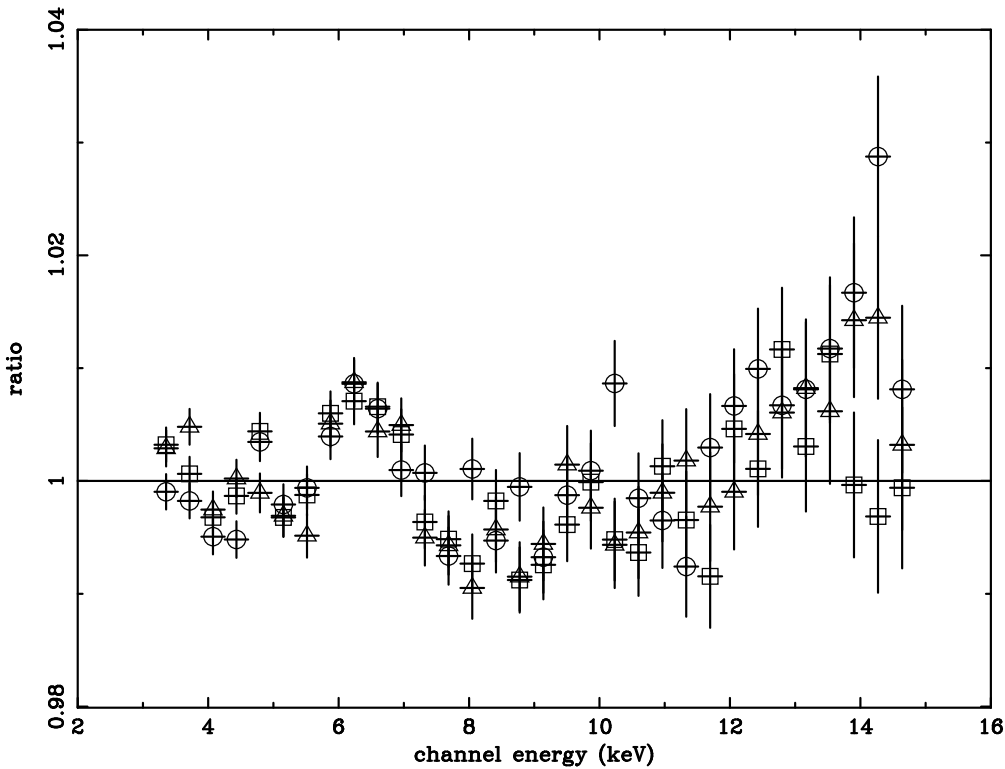}
\hfil}
}

\caption
{\footnotesize The ratio of the Crab nebula data to a best fit absorbed power law model, as
measured by RXTE PCA during the same time interval that A496 was
observed. Circle, square and triangle symbols correspond to data
taken on 15/12/98, 29/12/98 and 13/1/99, respectively.  
}
\end{figure}

\subsection{Thermal Models}

Table 1 shows all the models and their respective parameters 
used to fit the PCA data.  
For all the models reported, we fixed the 
Galactic hydrogen column density and the
redshift at $4.5 \times 10^{20} \, {\rm cm^{-2}}$ and
$0.033$, respectively. We report errors (90\% confidence limit) 
for models with a $\chi^2/\nu$ of less than 2. 

Model A is a single isothermal mekal model
(Mewe, Gronenschild, \& van den Oord 1985; Mewe, Lemen, \& van den
Oord 1986; Liedahl, Osterheld, \& Goldstein 1995).  The abundances of
all elements are tied together and are allowed to vary, except for that 
of He which is fixed at the Solar value.
However, this model is not satisfactory based on 
the $\chi^2$ statistic and the ratio of data to model shown in Figure~2a.
Allowing the Ni abundance to vary independently improves the quality 
of the fit.
Model B is therefore identical to model A except that we allowed the 
Ni abundance to vary
independently. This model is a significant improvement
over model A (Figure~2b).  The best fit Ni/Fe ratio in this model
is $\sim 3.6$, consistent with the central Ni/Fe measured by Dupke \& White
(2000) using ASCA. In all other models reported below, we fitted 
the data with both fixed and variable Ni/Fe ratio.
Since we consistently find that models with independently varying
Ni abundance 
(specifically, with Ni/Fe greater than one) produce the best
fits, we only discuss models with independent Ni abundances
hereafter.

Model B can be further improved. In particular, the cooling flow in A496
has been ignored. We construct model C by adding a 
cooling flow component to model B. We tie the upper temperature
of the cooling flow model to the temperature of the mekal component
and also tie the abundances of the two components together.  
Of the models considered thus far, model C 
produces the best fit to the single beam PCA data (Figure~2c).
The total unabsorbed 2--10~keV flux in this model is 
$7.1 \times 10^{-11}\, {\rm erg \, cm^{-2} \, s^{-1}}$. About
24\% of the emission in this band is from the cooling flow
component. From examination of archival A496 Ginga
data, we find a total flux of $7.3 \times 10^{-11}\, 
{\rm erg \, cm^{-2} \, s^{-1}}$ for this model. Hence, the RXTE
and Ginga calibrations are in relatively good agreement.  

\begin{figure}[tbh]
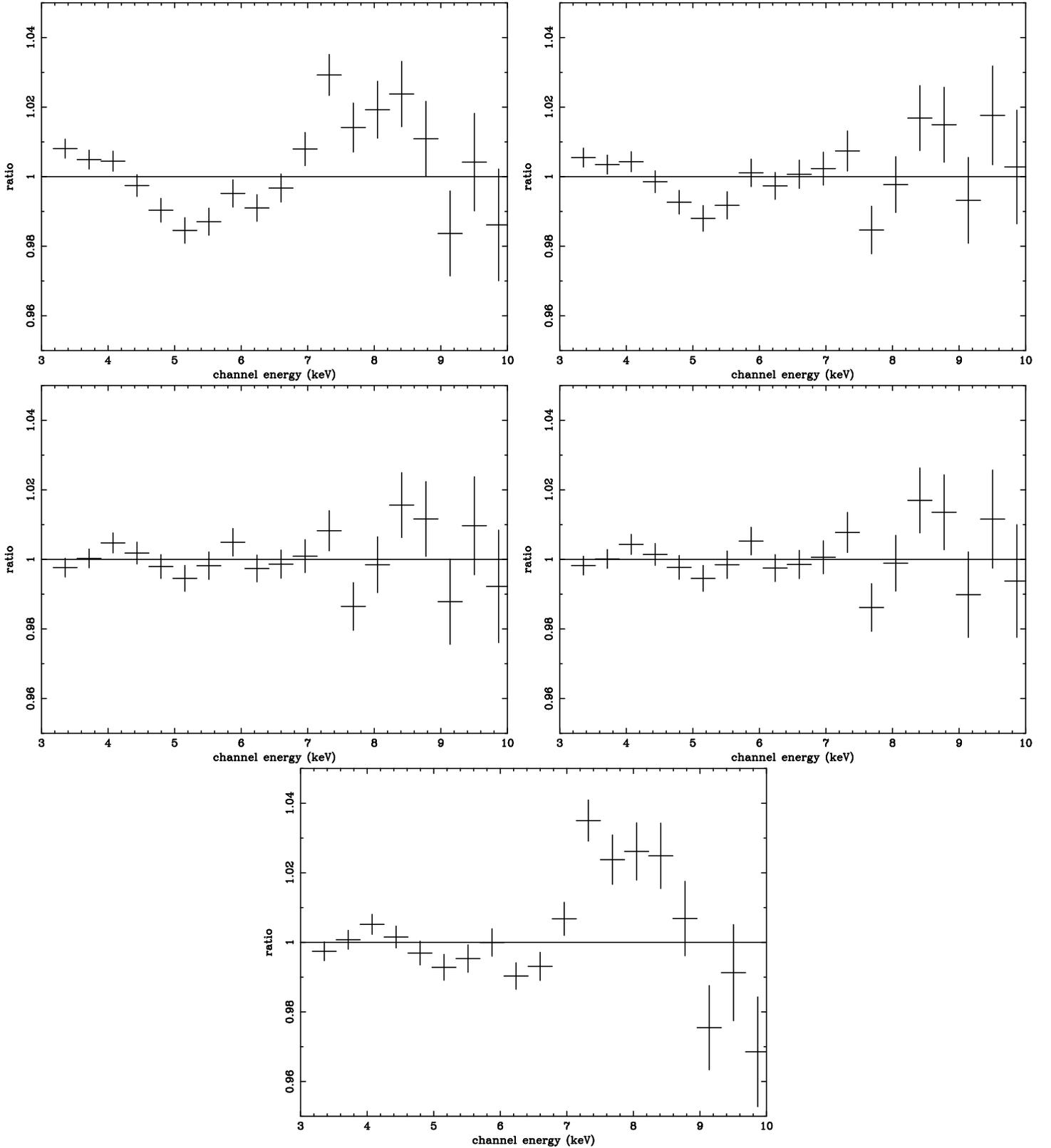

\centerline{
{\hfil \hfil
\psfig{figure=figure2a.ps,height=2.8truein,angle=270.0}
\psfig{figure=figure2b.ps,height=2.8truein,angle=270.0}
\hfil \hfil} }
\centerline{
{\hfil \hfil
\psfig{figure=figure2c.ps,height=2.8truein,angle=270.0}
\psfig{figure=figure2d.ps,height=2.8truein,angle=270.0}
\hfil \hfil} }
\centerline{
{\hfil
\psfig{figure=figure2e.ps,height=2.8truein,angle=270.0}
\hfil}
}
\caption{\footnotesize (From top left to right)
The ratio of A496 data taken with RXTE PCA to model for
model A (a), model B (b),
model C (c), model E (d), and model F (e).
}
\end{figure}

Model D consists of two mekal components with their abundances tied 
together. Although, statistically, this model provides a good fit to
the data, neither of the temperature components can be well
constrained. Therefore, we have not reported any parameter errors
for this model. 

Model E is composed of a multi-temperature mekal component with a
Gaussian temperature distribution. This model gives the best 
fit to the data (Figures 2d and 3).  The total unabsorbed 2--10~keV flux
for this model is $7.0 \times 10^{-11}\, {\rm erg \, cm^{-2} \, s^{-1}}$.
The best fit $\sigma_{kT}
\sim 0.9$~keV indicates that the temperature distribution is narrow. 
The variations in the ratio of data to model below
10~keV are within the systematic errors observed for the Crab (i.e. compare
Figures 1 and 2d). 

Finally, we fit the data using a cooling flow plus the polytropic gas 
distribution derived by Markevitch et al. (1999) from ASCA data. The 
polytropic component was approximated by calculating the
relative emission measures, derived using the Markevitch et al.
(1999) $\beta$-model fits to the A496 surface brightness profile,
in five temperature bins.
Table~2 shows the emission measure fraction and
emission-weighted temperature for each bin. 
Model F represents the temperature distribution for emission within
the ASCA field of view (fov) only (Figure~2e). 
Model G extends the emission to the edge of the
RXTE fov ($R \sim 2.4{h_{70}}^{-1}$ Mpc), holding the temperature
constant at the value corresponding to the edge of the ASCA fov.
Model H extrapolates the ASCA temperature profile using the polytropic
index and density slope derived by Markevitch et al. The cooling
flow component contributes 27\%, 34\%, and 36\% to the total 
emission in models F, G, and H, respectively.  
As seen from Table~1, none of these models fit the data well.
Removing the cooling flow component from these models produces even
poorer fits.  Evidently, models with a broad temperature distribution
that include substantial emission at relatively low temperatures are
inconsistent with the data.

\begin{figure}[h]
\centerline{
{\hfil
\psfig{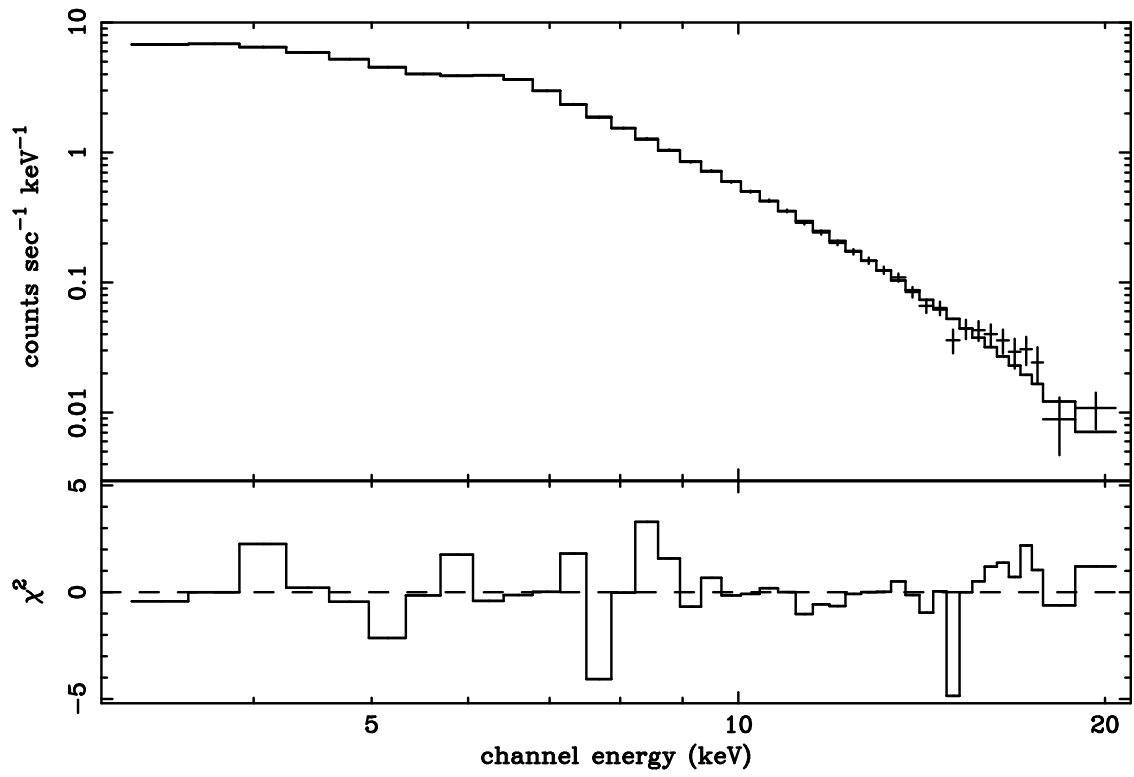}
\hfil}
}

\caption
{\footnotesize RXTE PCA data of A496 and folded model E in Table 1. $\chi^2$
is plotted in the lower panel.
}
\end{figure}

\subsection{Nonthermal Emission}

To search for evidence of hard X-ray nonthermal emission from A496, we
added a power law component to the best fit model E in Table~1. We
fixed the photon index of the power law at 1.8.  This is
because a nonthermal hard X-ray power law with $\Gamma=1.8$ is
detected in the BeppoSax hard X-ray spectrum of A2199 (Kaastra et
al. 1999), a nearby ($z \approx 0.031$) cluster that is nearly
identical in its X-ray properties to A496. Hence, we assume that the nonthermal
component, if it exists, has similar properties to that in A2199.
The $\chi^2/\nu$ for this model, however, decreases by 0.2 to $40.8/40$ which
represents a statistically insignificant improvement (see Table~1).  The
upper limit (90\% confidence limit) to the contribution of this
component in the 3-20~keV band is 6\%. The upper limit to the power
law flux at 20~keV is $1.1 \times 10^{-13} \, {\rm erg \, cm^{-2} \,
s^{-1} \, keV^{-1}}$.  The Cosmic X-ray background (CXB) fluctuations
are about 8\% rms of the mean CXB per RXTE fov (Valinia et
al. 1999). The amplitude of the fluctuations amounts to $6.5 \times
10^{-14} \, {\rm erg \, cm^{-2} \, s^{-1} \, keV^{-1}}$ per RXTE fov
at 20~keV.  Therefore, the upper limit quoted above corresponds to a
2 $\sigma$ fluctuation in the CXB.

For comparison we note that the non-thermal component in A2199
measured by Kaastra et al. (1999) using BeppoSAX has a flux at 20~keV
of $3.2 \times 10^{-13} \, {\rm erg \, cm^{-2} \, s^{-1} \, keV^{-1}}$.
The contribution of the non-thermal component in the similar 3-20~keV
band is about 12\%. 
A comparable hard X-ray tail in A496 would have been easily detected by RXTE.

Neither allowing the power-law slope to be a free parameter, nor
substituting Model C for Model E as the thermal model, significantly
improves the quality of the fit or increases the
upper limit to the non-thermal contribution.

\section{DISCUSSION and CONCLUSIONS}

We have presented the results of our analysis of the highest signal-to-noise
hard X-ray observations of A496.
The large effective area and well calibrated response of RXTE enables us to
place strong constraints on the distribution function of temperature
in the total cluster volume. The distribution peaks at a temperature
of $\approx 4.1~keV$; however cooler emission is detected and can be
modeled as an additional isothermal component or as a cooling flow
($\dot{M} \sim 285 \, {\rm M_{\odot} \, yr^{-1}}$). The data is also
adequately modeled by a Gaussian distribution with $\sigma_{kT}=$0.9~keV.
Since a mass flow rate of $\sim 285 \, {\rm M_{\odot} \, yr^{-1}}$ is
high compared to Einstein, ROSAT and ASCA measurements ($\sim 100 \, {\rm
M_{\odot} \, yr^{-1}}$, White et al. 1997; 
Peres et al. 1998; Allen et al. 2000) it is
likely that the true model is a combination of a cooling flow and
additional multi-temperature plasma.
In principle, this can set limits on the degree of homogeneity of 
cluster gas (Allen et al. 1992). 
However, the temperature profile derived from ASCA data by Markevitch
et al. (1999) produces a broad, flat distribution and is ruled out
unless there is a temperature inversion outside the ASCA fov.
However, no temperature inversion was found outside the half
virial radius (roughly ASCA field of view in the case of A496) 
for clusters with $z>0.03$ in the Santa Barbara cluster simulation
project (Frenk et al. 1999). 
These temperature gradients have been questioned (e.g., Irwin,
Bregman, \& Evrard 1999; Kikuchi et al. 1999) and await {\it
XMM-NEWTON} observations for confirmation.

Metal abundances of 1/3 solar and a Ni/Fe ratio of 3.6 are found for this
cluster. Both the peak in the temperature distribution 
and the metallicities are in excellent agreement  
with previous measurements of this cluster with ASCA (Mushotzky et al. 1996). 
The Ni/Fe ratio is consistent with that measured in the central
region of the cluster (Dupke \& White 2000). If the apparently
high Ni/Fe ratio
observed in the center of the cluster had been due to depressed
emission from a resonantly-scattered Fe K$\alpha$ line then 
a lower measured Ni/Fe ratio would have been measured by
RXTE since it collects
all the Fe K$\alpha$ photons, whether or not they are scattered. So,
we conclude that the measured Ni/Fe ratio is a true measurement of the relative
abundances of these two elements in the central region 
and is not due to optical depth effects.
Since Ni is mostly originated in type I SNe, the ratio of Ni/Fe
sets constraints on the production of Fe in type I SNe 
throughout the cluster. 

No significant hard X-ray non-thermal emission was detected. Hard X-ray
emission in clusters of galaxies is expected from detailed models of
intracluster medium (ICM) evolution. The contributing processes are
expected to be inverse Compton (IC) scattering, synchrotron radiation,
nonthermal bremsstrahlung emission and Coulomb losses to the ICM
(e.g. Sarazin 1999). Since radiation fields (e.g. optical/IR and
X-ray) and magnetic fields ($\lesssim 1 \, \mu$G) are low in the ICM,
highly relativistic electrons mainly lose energy via IC scattering off
cosmic microwave background photons (Sarazin \& Lieu 1998).  For an
electron of energy 20~keV, its lifetime is dominated by IC losses
given by \bq t_{IC}={\gamma m_e c^2 \over {4 \over 3} \sigma_T c
\gamma^2 U_{CMB}} \approx 4.8 \times 10^8 ({\gamma \over 4800})^{-1}\,
{\rm yr}, \eq where $\gamma$, the Lorentz factor, is about
4800. Clearly, these particles have very short lifetimes and are
present only in clusters where there has been current or very
recent particle acceleration such as in ICM shocks or cluster mergers.
In this context, the detection of nonthermal emission from A2199 with
BeppoSax is surprising.  A more plausible mechanism for producing hard
tails in clusters without radio halos is non-thermal bremsstrahlung by
suprathermal (but subrelativistic) electrons (Kempner \& Sarazin 2000,
Blasi 2000). Ongoing turbulent particle acceleration is still required
to overcome Coulomb losses; however, the power requirements may be
acceptable -- even for relatively relaxed clusters.  A2199 has very
similar properties to A496 -- it is at a redshift of 0.032 with an
average cluster temperature of 4.2~keV, and it is therefore somewhat
surprising that there may be a hard tail in the former but not the latter.
We caution that the existence of nonthermal hard X-ray emission from
clusters of galaxies is not experimentally established as a general
phenomenon. To date, 
positive detection of this component has been reported only for the Coma
cluster
(Rephaeli, Gruber, \& Blanco 1999; Fusco-Femiano et al. 1999), A2256
(Fusco-Femiano et al. 2000) and A2199. However, Coma and A2256 are 
very dynamically active clusters undergoing merging while A2199 is not. 
A complete hard X-ray survey of galaxy clusters is needed to establish 
whether the presence of this component is common or exceptional.

\acknowledgements

We thank Keith Jahoda for providing the Crab data. This research has
taken advantage of the LEDAS archival database. 

%

\clearpage
\begin{deluxetable}{cccccccl}
\tablecolumns{8}
\tablecaption{Models and Best Fit Parameters}
\tablewidth{0pc}
\tablehead{ Model &  $kT$ (keV)  & 
$\sigma_{kT}$ (keV)   &  Abundances  &  Ni &  Ni/Fe  &
$\dot{M}$ (${\rm M_{\odot} \, yr^{-1}}$)  &  $\chi^2/\nu$\tablenotemark{*} \\
  &  &  &  ($\times$ solar) &  &  &  &   }
\startdata
A  &   3.99  &  -- &  0.32  & 0.32 &  1 &  -- &  129.5/43 \nl
B  &   $3.93\pm0.02$  &  -- &  $0.32\pm0.01$  & $1.17\pm0.18$ & 3.7 &  --  &  
69.7/42  \nl
C  &   $4.14^{+0.10}_{-0.08}$  &  --  & $0.33\pm0.01$  &  $1.15\pm0.19$ &  
3.5  &   $285\pm88$  &   41.7/40  \nl
D  &   3.5 \&  6.8 &  --  &  0.34 &  1.26 & 3.7  & --  &   41.0/39  \nl
E  &   $3.81^{+0.05}_{-0.06}$  &  $0.90\pm0.14$  &  $0.34\pm0.01$  & 
$1.21\pm0.19$  &  3.6  &   --  &   41.0/41  \nl
F  &   --  &  -- &  0.22 &  0.29  & 1.3  &  315  &  138.4/40  \nl
G  &   --  &  --  &  0.27  &  0.33 & 1.2  &  270  &  292.5/40  \nl
H  &   --  &  --  &  0.25  &  0.37  & 1.5  &  295  &  503.4/40  \nl
\enddata
\tablenotetext{*}{chi-squared over degrees of freedom} 
\end{deluxetable}

\begin{deluxetable}{clll}
\tablecolumns{4}
\tablecaption{Emission Measure Percentages Assuming ASCA
T-Profile\tablenotemark{*}}
\tablewidth{0pc}
\tablehead{ Temperature Bin (keV) &  Model F  & Model G & Model H }

\startdata
5.2--6.2 &  18(5.6)   &     10(5.6)   &       10(5.6)  \nl
4.2--5.2  &   27(4.7)  &    18(4.7)   &        18(4.7)  \nl
3.2--4.2  &   42(3.7)  &    22(3.7)   &        22(3.7)  \nl
2.2--3.2  &   13(3.0)  &    50(3.0)   &       30(2.7)  \nl
$<2.2$   &   0        &     0        &       20(2.0)   \nl
\enddata
\tablenotetext{*}{emission-averaged temperature (keV) in each bin in
parenthesis}
\end{deluxetable}

\clearpage

\end{document}